\documentclass[conference]{IEEEtran}
\IEEEoverridecommandlockouts

\usepackage{cite}
\usepackage{amsmath,amssymb,amsfonts}
\usepackage{graphicx}
\usepackage{textcomp}
\usepackage{xcolor}
\usepackage{booktabs}
\usepackage{array}
\usepackage{multirow}
\usepackage{url}
\usepackage{balance}

\newcommand{\method}{QRC-Agent}
\newcommand{\hybrid}{LLM-Hybrid}
\newcommand{\best}{best-so-far}
\def\BibTeX{{\rm B\kern-.05em{\sc i\kern-.025em b}\kern-.08em
    T\kern-.1667em\lower.7ex\hbox{E}\kern-.125emX}}
\begin{document}
\title{\method: Hybrid LLM-Guided Search for Quantum Reservoir Architecture Design}

\author{
\IEEEauthorblockN{Krishna Bhatia}
\IEEEauthorblockA{\textit{QuantumAI Lab, Fractal Analytics} \\
India \\
krishna.bhatia@fractal.ai}
\and
\IEEEauthorblockN{Gautami Sanjay Naik}
\IEEEauthorblockA{\textit{Indian Institute of Information Technology Dharwad} \\
India \\
gautaminaik2000@gmail.com}
}

\maketitle

\maketitle

\begin{abstract}
Quantum reservoir computing (QRC) uses fixed quantum dynamics as a high-dimensional temporal feature map and trains only a lightweight classical readout. QRC is attractive for near-term quantum machine learning, but its performance depends strongly on architecture choices such as input encoding, reservoir depth, entanglement topology, measurement features, state-reset policy, feature construction, and readout regularization. We introduce \method, a simulator-based benchmark that formulates QRC design as constrained black-box architecture search and evaluates whether large language models can act as proposal controllers for this search problem. The benchmark compares five policies under identical evaluation budgets: random search, evolutionary search, Bayesian/TPE optimization, a feedback-based LLM agent, and \hybrid, which combines LLM proposals with memory, mutation, crossover, duplicate avoidance, and exploration. On NARMA10, Mackey-Glass forecasting, and temporal parity, \hybrid{} is the most consistent policy: it ranks first on NARMA10 and temporal parity and second on Mackey-Glass, narrowly behind evolutionary search. Under a 25-evaluation budget and three seeds, \hybrid{} improves over random search on all tasks, including a 23.6\% relative reduction in Mackey-Glass error. The results do not show that LLMs are universal QRC optimizers; rather, they show that generative models can be useful high-level controllers when embedded inside validated, reproducible hybrid search loops.
\end{abstract}


\section{Introduction}
Quantum reservoir computing (QRC) uses a fixed quantum system as a nonlinear dynamical feature map for temporal data. At each time step, an input is encoded into a quantum state or circuit, the reservoir evolves, observables are measured, and only a classical readout is trained. Because the quantum subsystem is not optimized by gradients, QRC is appealing for noisy intermediate-scale quantum settings where variational training can be expensive or unstable. Foundational QRC work showed that disordered quantum dynamics can support temporal machine learning~\cite{fujii2017harnessing}, while later studies improved computational capacity through spatial multiplexing~\cite{nakajima2019boosting} and explored natural QRC on superconducting devices~\cite{suzuki2022natural}.

However, QRC performance is highly configuration-dependent. A practitioner must choose how to encode inputs, how many qubits or virtual nodes to use, what circuit family or physical dynamics to instantiate, which observables to expose, whether to reset or continuously evolve the state, how much history to feed into the readout, and how to regularize the readout. Prior work has therefore begun to treat QRC design as a search problem, including memory-reset optimization~\cite{molteni2022memory}, genetic configuration for multi-task learning~\cite{xia2023configured}, and reproducible software tooling for QRC experiments~\cite{santos2026qrclab}. In parallel, LLM agents are increasingly used to generate, repair, and optimize quantum programs and circuits~\cite{vishwakarma2024qiskit,jern2025agentq,fu2025qagent}. These trends motivate the question studied here: can LLMs act as competitive search controllers for QRC architecture design?

We do not claim quantum advantage, hardware performance, or autonomous physical discovery. Instead, we study a narrower and auditable question: can an LLM serve as a useful proposal controller for QRC architecture search when every proposed design is schema-validated and scored by the same simulator? This framing is important for generative AI in quantum workflows because the generated object is not free-form code, but a constrained architecture proposal that must survive validation and task-level evaluation. Our contributions are: (1) a compact QRC architecture-search space with schema-validated configurations; (2) \method, a closed-loop benchmark that treats LLM guidance as a proposal policy rather than as a simulator or oracle; (3) \hybrid, which combines LLM proposals with classical search operators and memory; and (4) a matched-budget comparison against random, evolutionary, and Bayesian/TPE search on three temporal-learning benchmarks.

\section{Background and Related Work}
\subsection{Reservoir Computing and QRC}
Classical reservoir computing originated from echo-state networks and liquid-state machines, which use fixed recurrent dynamics and train only a simple readout~\cite{jaeger2001echo,maass2002real,lukosevicius2009reservoir}. QRC follows the same separation of fixed dynamics and trained readout, but replaces the reservoir with a quantum dynamical system. Given an input sequence $u_t$, the reservoir produces a feature vector $r_t$ through quantum evolution and measurement, and a ridge-regression readout predicts targets $y_t$:
\begin{equation}
    \hat{y}_t = W r_t + b, \quad
    W = \arg\min_W \sum_t \|y_t - W r_t\|_2^2 + \lambda \|W\|_2^2 .
\end{equation}
For regression tasks we minimize normalized root mean squared error (NRMSE); for temporal parity we minimize classification error, so lower scores are better for all reported tasks.

QRC studies have shown that quantum reservoirs can support nonlinear temporal processing, memory, and forecasting~\cite{fujii2017harnessing,nakajima2019boosting,suzuki2022natural}. At the same time, the field increasingly recognizes that reservoir performance is governed by trade-offs among memory retention, nonlinearity, measurement choice, dissipation, and reset policy~\cite{molteni2022memory,santos2026qrclab}. This makes QRC a natural target for architecture search: even when the quantum dynamics are not trained, many design choices remain.

\subsection{Search and LLM Agents for Quantum Workflows}
Non-LLM QRC configuration methods already exist. Configured QRC uses evolutionary search to tune reservoir dynamics for multi-task learning~\cite{xia2023configured}; Bayesian optimization is also a standard black-box optimizer for expensive model evaluations, with Optuna providing a practical TPE implementation~\cite{akiba2019optuna}. The novelty of this work is therefore not ``searching QRC'' in isolation. Instead, we benchmark whether LLM-guided controllers can provide sample-efficient, task-aware proposals under the same evaluation budget as classical search baselines. This distinction matters because an LLM can appear useful in anecdotal quantum-programming examples while still failing as an optimizer when compared against random, evolutionary, or Bayesian search.

Recent quantum-AI work suggests that LLMs can assist quantum software workflows. Qiskit HumanEval benchmarks LLMs on executable quantum-code generation tasks~\cite{vishwakarma2024qiskit}; Agent-Q fine-tunes LLMs for quantum circuit generation~\cite{jern2025agentq}; and QAgent uses multi-agent planning and reflection for OpenQASM programming~\cite{fu2025qagent}. Our setting is different: the target is not syntactic program correctness, but task-level predictive performance of a QRC architecture. We therefore treat the LLM as a proposal policy inside a validated black-box optimization loop, not as a simulator or oracle.

\section{Problem Formulation and Search Space}
We cast QRC design as black-box optimization over a constrained configuration space $\mathcal{C}$. Given task $\tau$ and evaluator $E$, the goal is
\begin{equation}
    c^* = \arg\min_{c \in \mathcal{C}} E(c,\tau),
\end{equation}
where $E$ simulates the reservoir, trains the readout after a washout period, and returns a held-out test score. Table~\ref{tab:space} summarizes the search space. The additional state, feature, and lag options are included because they are natural QRC/readout design choices rather than LLM-specific advantages; all methods search the same space.

\begin{table}[t]
\caption{QRC architecture search space used by all methods.}
\label{tab:space}
\centering
\scriptsize
\begin{tabular}{ll}
\toprule
Field & Values \\
\midrule
Qubits & $3$--$8$ \\
Depth & $1$--$12$ \\
Encoding & $R_x$, $R_y$, $R_z$, $R_xR_y$, $R_yR_z$ \\
Topology & none, chain, ring, all-to-all, random \\
Entangler & CX, CZ \\
Measurement & all $Z_i$; all $Z_i$ plus pairwise $Z_iZ_j$ \\
Input scale & $0.5,1,2,\pi$ \\
Reservoir scale & $0.25,0.5,1,2$ \\
State mode & reset, continuous \\
Input reupload & $1,2$ \\
Feature mode & raw, quadratic, delta, quadratic+delta \\
Readout lag & $0,1,2,3$ \\
Ridge $\alpha$ & $10^{-5},10^{-4},10^{-3},10^{-2},10^{-1},1$ \\
\bottomrule
\end{tabular}
\end{table}

\section{Method}
The unit of comparison is one completed QRC evaluation, not one controller step. Thus, a trial consumes budget only when a valid configuration is evaluated by the common simulator. The LLM is never used to estimate scores, modify the simulator, alter train/test splits, or access held-out labels. It receives only the task name, the configuration schema, and a compact history of previous evaluated configurations and scores for its own search loop. This protocol separates generative proposal quality from objective evaluation and makes the LLM-guided policies directly comparable to non-LLM baselines.

\begin{figure}[t]
    \centering
    \includegraphics[width=0.83\linewidth]{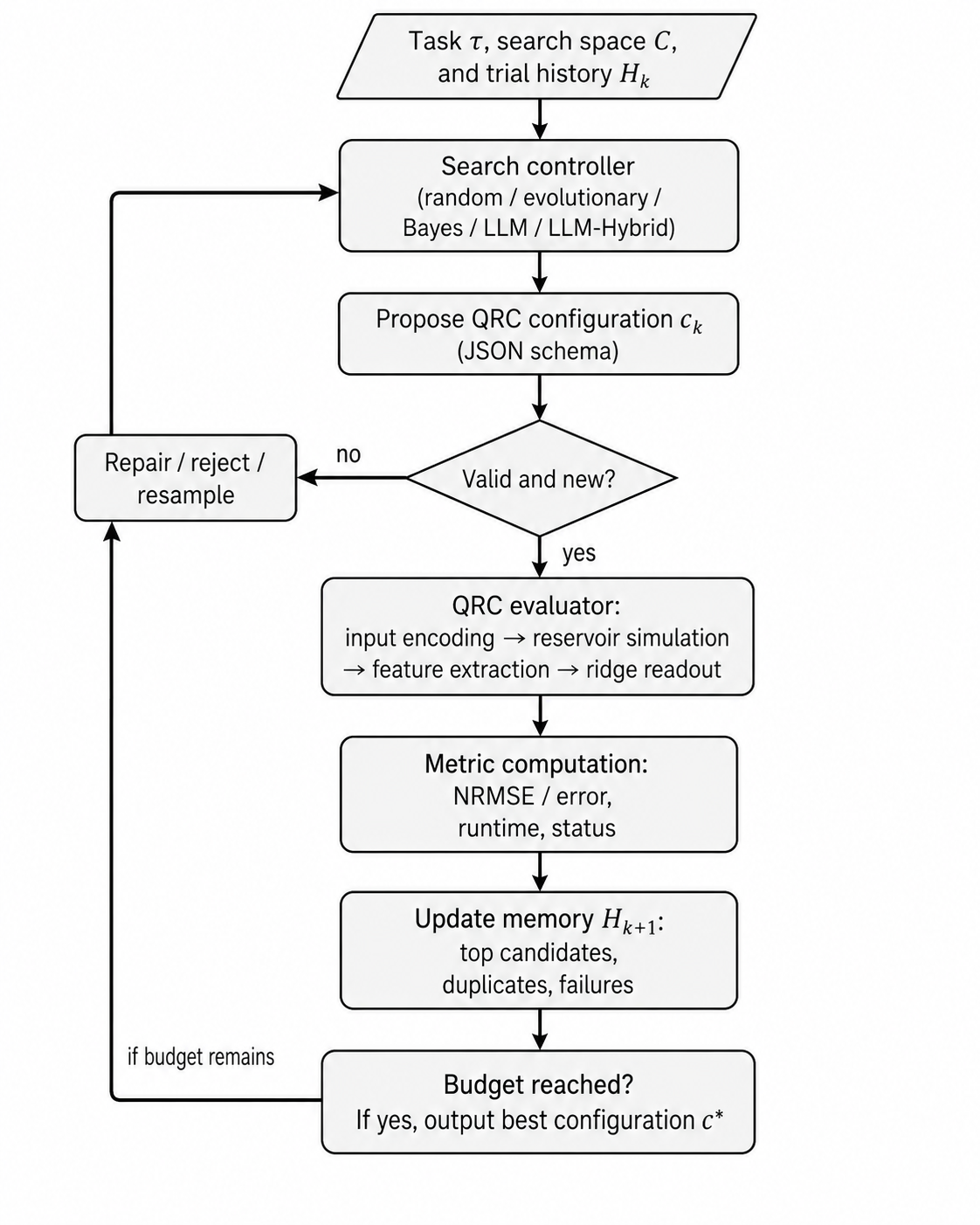}
    \caption{Closed-loop QRC architecture-search workflow. A controller proposes schema-valid QRC configurations, the evaluator scores each candidate, and score feedback updates the search memory.}
    \label{fig:loop}
\end{figure}

We compare five policies. \textbf{Random search} samples uniformly from the constrained space and serves as a minimum-effort baseline. \textbf{Evolutionary search} keeps high-performing configurations and mutates architecture fields, representing a strong non-LLM heuristic baseline. \textbf{Bayesian/TPE} uses Optuna's tree-structured Parzen estimator~\cite{akiba2019optuna}, representing a standard black-box optimization baseline. \textbf{LLM-Agent} prompts an LLM with the task, schema, and compact trial history, and asks for a valid JSON proposal. \textbf{\hybrid} extends LLM-Agent with elite memory, duplicate avoidance, task-specific priors, local mutation around strong candidates, crossover between elite candidates, and random exploration. The comparison therefore tests two questions: whether LLM proposals are competitive with classical search, and whether classical search machinery stabilizes LLM-guided optimization.

\section{Experiments}
We evaluate on three standard temporal-learning benchmarks. \textbf{NARMA10} tests nonlinear memory; \textbf{Mackey-Glass} tests chaotic time-series forecasting; and \textbf{temporal parity} tests nonlinear memory over binary sequences. Each method is run for 25 QRC evaluations on three seeds. Scores are averaged over seeds, and lower is better for all tasks. The main question is not whether LLMs dominate every baseline, but whether LLM guidance improves the reliability and sample efficiency of QRC architecture search.

\begin{table}[t]
\caption{Final-budget results with 25 evaluations and three seeds. Lower is better. Rank is computed within each task.}
\label{tab:results}
\centering
\scriptsize
\begin{tabular}{llcccc}
\toprule
Task & Method & Mean & Std. & Best & Rank \\
\midrule
\multirow{5}{*}{Mackey-Glass}
& Evolutionary & \textbf{0.020138} & 0.003143 & 0.017365 & 1 \\
& \hybrid & 0.020775 & 0.005679 & \textbf{0.015275} & 2 \\
& Bayes/TPE & 0.021571 & 0.007173 & 0.015470 & 3 \\
& LLM-Agent & 0.022009 & 0.002227 & 0.019497 & 4 \\
& Random & 0.027176 & 0.012631 & 0.016616 & 5 \\
\midrule
\multirow{5}{*}{NARMA10}
& \hybrid & \textbf{0.835414} & 0.061058 & 0.781304 & 1 \\
& LLM-Agent & 0.839881 & 0.063029 & 0.782979 & 2 \\
& Random & 0.846353 & 0.063220 & 0.787484 & 3 \\
& Bayes/TPE & 0.892192 & 0.099721 & 0.789376 & 4 \\
& Evolutionary & 0.895465 & 0.096128 & 0.791715 & 5 \\
\midrule
\multirow{5}{*}{Temporal parity}
& \hybrid & \textbf{0.430000} & 0.020000 & 0.410000 & 1 \\
& LLM-Agent & 0.433333 & 0.046458 & \textbf{0.380000} & 2 \\
& Random & 0.441667 & 0.017559 & 0.425000 & 3 \\
& Bayes/TPE & 0.445000 & 0.000000 & 0.445000 & 4 \\
& Evolutionary & 0.451667 & 0.005774 & 0.445000 & 5 \\
\bottomrule
\end{tabular}
\end{table}

\begin{table}[t]
\caption{Aggregate consistency across the three tasks. Average rank is computed from Table~\ref{tab:results}; lower is better.}
\label{tab:consistency}
\centering
\scriptsize
\begin{tabular}{lcc}
\toprule
Method & Avg. rank & Tasks better than random \\
\midrule
\hybrid & \textbf{1.33} & \textbf{3/3} \\
LLM-Agent & 2.67 & 3/3 \\
Evolutionary & 3.67 & 1/3 \\
Bayes/TPE & 3.67 & 1/3 \\
Random & 3.67 & -- \\
\bottomrule
\end{tabular}
\end{table}

\begin{figure}[t]
    \centering
    \includegraphics[width=0.92\linewidth]{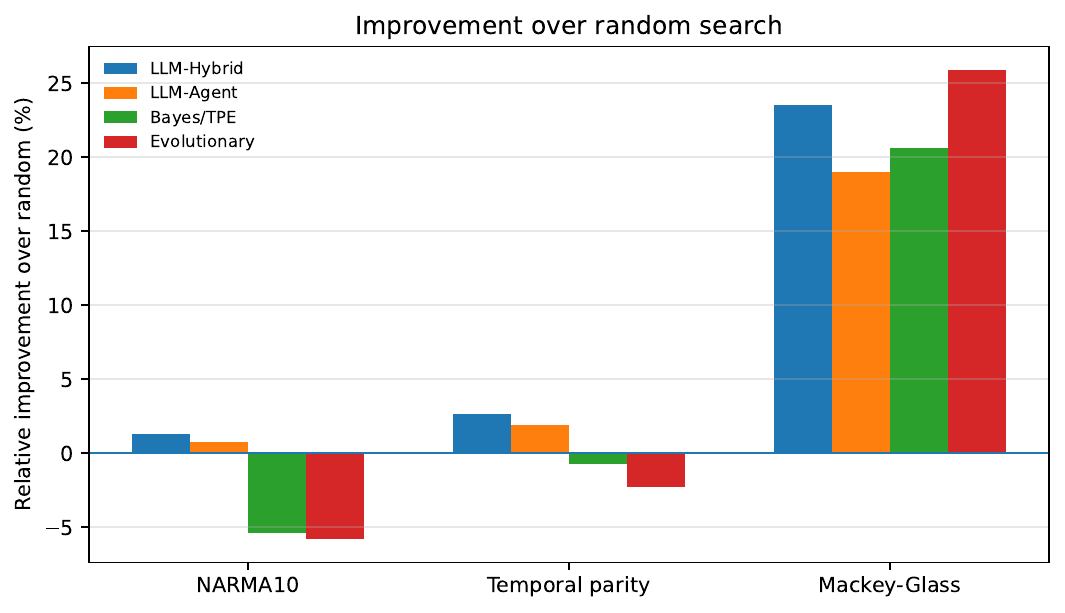}
    \caption{Relative improvement over random search under the same evaluation budget. \hybrid{} improves over random search on all tasks, with the largest gain on Mackey-Glass.}
    \label{fig:improvement}
\end{figure}

Table~\ref{tab:results}, Table~\ref{tab:consistency}, and Fig.~\ref{fig:improvement} show that \hybrid{} is the most consistent method across tasks. It ranks first on NARMA10 and temporal parity, and second on Mackey-Glass, narrowly trailing evolutionary search in mean score while achieving the best single-seed score on that task. Compared with random search, \hybrid{} reduces NARMA10 error from 0.846353 to 0.835414, temporal-parity error from 0.441667 to 0.430000, and Mackey-Glass error from 0.027176 to 0.020775. These correspond to relative error reductions of 1.3\%, 2.6\%, and 23.6\%, respectively. \hybrid{} also improves over the plain LLM-Agent mean score on all three tasks, with relative reductions of 0.5\% on NARMA10, 0.8\% on temporal parity, and 5.6\% on Mackey-Glass.

The strongest conclusion is therefore not that an LLM alone dominates classical search. Evolutionary search remains best on Mackey-Glass in mean performance, and the margins on NARMA10 and temporal parity are modest. Rather, the evidence supports a narrower claim: LLM proposals become more reliable when embedded inside a classical optimization scaffold. In this setting, the LLM contributes structured, task-aware proposals, while memory, mutation, crossover, duplicate avoidance, and random exploration reduce brittleness.


\section{Discussion and Limitations}
\method{} is deliberately modest and auditable. The LLM is constrained to propose schema-valid configurations, and every score is produced by the same simulator and readout pipeline. This makes it possible to compare LLM-guided search directly against non-LLM optimizers rather than relying on anecdotal examples of generated circuits. The result is most relevant to generative AI for quantum workflows: the LLM is useful not because it replaces the quantum evaluator, but because it can steer a constrained proposal process that remains externally validated.

Several limitations remain. First, all experiments are simulator-based and use small reservoirs, so the results are not evidence of quantum advantage or hardware performance. Second, the benchmark uses three seeds and a fixed 25-evaluation budget; larger sweeps may change rankings or reduce variance. Third, \hybrid{} contains hand-designed operators, so its improvement over LLM-Agent should be interpreted as evidence for hybrid LLM-guided search rather than for the LLM alone. Fourth, the search space is restricted to gate-based reservoirs and simple measurement/readout features. Future work should add noise models, hardware connectivity constraints, larger budgets, physical QRC reservoirs, and ablations isolating elite memory, mutation, crossover, task priors, and exploration.

\section{Artifact and Reproducibility}
The implementation logs every trial as a JSONL record containing the method, task, seed, configuration, score, runtime, and validation status. The code separates task generation, QRC simulation, search policies, prompt construction, and result analysis. The review artifact will include the search-space definition, exact prompts, random seeds, raw JSONL logs, plotting scripts, and analysis scripts used to generate Table~\ref{tab:results}, Table~\ref{tab:consistency}, and Fig.~\ref{fig:improvement}. This structure allows reviewers to reconstruct \best{} curves, inspect repeated or invalid configurations, verify matched evaluation budgets, and reproduce the reported aggregate rankings.

\section{Conclusion}
We presented \method, a benchmark for LLM-guided QRC architecture search under constrained, matched-budget black-box optimization. Across NARMA10, Mackey-Glass, and temporal parity, \hybrid{} achieved the best average rank, improved over random search on all tasks, and improved over the plain LLM-Agent on all mean scores. The results are intentionally modest: they do not establish quantum advantage or show that LLMs replace classical optimizers. They instead support a practical design pattern for generative AI in quantum workflows: use the LLM as a structured proposal controller, but keep validation, memory, exploration, and objective evaluation outside the model.
\balance
\bibliographystyle{IEEEtran}
\bibliography{references}

\end{document}